\documentclass[%
 reprint,
 amsmath,amssymb,
 aps,
]{revtex4-1}

\usepackage{graphicx}
\usepackage[bb=boondox]{mathalfa}
\usepackage{dcolumn}
\usepackage{bm}
\usepackage{subcaption}
\usepackage{amsmath}
\usepackage{mathtools}
\usepackage{physics}
\usepackage{comment}
\usepackage[normalem]{ulem}
\usepackage{tikz-network}
\usepackage{ragged2e} 
\usepackage{comment}
\usepackage{todonotes}
\usetikzlibrary{decorations.pathmorphing}
\DeclareCaptionJustification{justified}{\justifying}

\usepackage{hyperref}
\hypersetup{
	colorlinks=true,
	linkcolor=blue,
	citecolor=blue,
	filecolor=magenta,
	urlcolor=cyan
}

\begin{document}

\title{An unbiased measure over the matrix product state manifold}

\def\UCL{{London Centre for Nanotechnology, University College London,
Gordon St., London WC1H 0AH, United Kingdom}}

\author{Sebastian Leontica}
\email{sebastian.leontica.22@ucl.ac.uk}
\affiliation{\UCL}
\author{Andrew G. Green}
\affiliation{\UCL}

\date{\today}
\UseRawInputEncoding

\begin{abstract}
Matrix product states are useful representations for a large variety of naturally occurring quantum states. Studying their typical properties is important for understanding universal behavior, including quantum chaos and thermalization, as well as the limits of classical simulations of quantum devices. We show that the usual ensemble of sequentially generated random matrix product states (RMPS) using local Haar random unitaries is not uniform when viewed as a restriction of the full Hilbert space. As a result, the entanglement across the chain exhibits an anomalous asymmetry under spatial inversion. We show how to construct an unbiased measure starting from the left-canonical form and design a Metropolis algorithm for sampling random states. Some properties of this new ensemble are investigated both analytically and numerically, such as the resulting resolution of identity over matrix product states and the typical entanglement spectrum, which is found to differ from the sequentially generated case.
\end{abstract}

\maketitle

\textit{Introduction}---
Since their discovery as an accurate representation of zero temperature states of a quantum system in 1D via the density matrix renormalization group algorithm \cite{White1992, White1993, Schollwock2005, Schollwock2011}, matrix product states (and tensor networks more broadly) have become some of the most widely adopted tools in computational science. Beyond their invaluable use in condensed matter physics \cite{Cirac2009, Orus2014,Verstraete2008}, tensor networks are also routinely used in computational chemistry calculations \cite{White1999,Baiardi2020, Szalay2015}, simulations of quantum algorithms \cite{Markov2008, Pan2022}, machine learning and artificial intelligence \cite{Stoudenmire2016, Rieser2023}, and optimization \cite{Hao2022, Akshay2024}.


The defining feature of tensor networks is their ability to efficiently represent states of many-body systems which are constrained to have only low correlations. Matrix product states (MPS), in particular, cover the space of area-law states, which are characterized by a finite amount of entanglement across a cut and an exponential decay of correlations with distance in the chain. Seen as an ensemble, they represent a very different class from fully random states, which typically have almost maximal entanglement. They are often considered a more realistic collection of states for modeling quantum systems, and can show up in a variety of settings such as systems with local interactions, shallow quantum circuits, disordered systems or monitored dynamics that leads to decoherence over large scales \cite{Verstraete2006, Skinner2019, Li2019, Smith2024, Doggen2021, Serbyn2021}.

Ensembles of random tensor networks have been proposed for studying typical entanglement properties in constrained many-body Hilbert subspaces. They have been explored in several works, leading to important results about entanglement in holographic quantum gravity \cite{Chirco2018, Vasseur2019} and quantum chaos \cite{Leontica2025, Hallam2019}. They were also used to classically simulate the operation of quantum computers, most prominently by sampling from random quantum circuits with a fidelity comparable to state of the art experiments \cite{Arute2019, Zhou2020, Ayral2023}. A range of results have been established to understand their properties, such as typical expectation values of local observables and entanglement \cite{Garnerone2010(2),Garnerone2010,Hakerkamp2021}, expected correlation length \cite{Haag2023, Lancien2021}, nonstabilizerness \cite{Chen2024}, and anticoncentration \cite{Lami2025, Sauliere2025}. The most common paradigm to produce an ensemble of non-translationally invariant random matrix product states is to assume independent unitary matrices of dimension $dD$ are placed at each site, each drawn from the Haar distribution of the unitary group. This is a natural assumption if one constructs the MPS on a quantum computer sequentially by applying unitaries on neighboring sites according to a ladder pattern. The approach is straightforward and leads to efficient sampling and theoretical results via the replica approach, but it produces a biased distribution with respect to the Fubini-Study metric on the matrix product state manifold, induced by its embedding into the full Hilbert space. This gives rise to a number of undesirable features, such as an observed asymmetry under flipping the ends of the chain, which we illustrate in Fig. \ref{fig:ent_chain} below. We can trace this odd behavior to the choice of working with left-canonical MPS, which induces a preferred orientation. In principle, one could artificially resolve the asymmetry by shifting the center of orthogonality to the middle of the chain, but we argue that this is not a good solution, as we are left with an artificial discontinuity in entanglement. Several applications implicitly assume that an unbiased ensemble of states is used \cite{Green2016, Leontica2025, Azad2023}, so neglecting this aspect may lead to misleading statistical results. 

Based on recent advances in our understanding of the geometry of matrix product state manifolds \cite{Haegeman2013}, we show how to correct the product Haar measure in order to recover an unbiased measure. We begin the investigation of this ensemble by showing that it can be used to produce a resolution of identity over matrix product states. This is an important feature, as it allows the use of the MPS manifold as a semi-classical phase space for path integral constructions \cite{Green2016,Leontica2024}. Previous theoretical analysis based on the Weingarten calculus is still feasible, though significantly more difficult for this ensemble, due to the complexity of expressing determinants as polynomials in the entries of a matrix. Arguments related to the typical correlation length \cite{Movassagh2022} also fail to hold, as the transfer matrices involved in correlation calculations are no longer sampled independently. We use free probability to derive an approximate form for the typical entanglement spectrum and show that its limiting behavior differs from that of RMPS at large bond dimension.

The volume factors introduced by the correction couple the distributions of the unitaries located at different sites, making sampling non-trivial. We design an implementation of the Metropolis-Hastings algorithm to produce samples from the unbiased distribution and numerically test our results.

\textit{Matrix product states}---We start by giving a brief overview of the construction of matrix product states and the geometry of the manifold they live on. For a system with $N$ sites and open boundary conditions, which we assume throughout this work, a matrix product state (MPS) is defined via the local 3-tensors $\mathcal{A} = \{A^{[i]}\}_{i=1}^N$ with a physical index of dimension $d$ and two bond indices going up to the bond dimension $D$. The bond indices are contracted sequentially to produce the matrix product state
\begin{equation}
\begin{split}
    \ket{\Psi[\mathcal{A}]} &= \sum_{\{n_i\}} A^{[1]}_{n_1}A^{[2]}_{n_2} \dots A^{[N]}_{n_N} \ket{n_1 n_2 \dots n_N}, \\
    &= \begin{tikzpicture}[baseline={([yshift=1ex]current bounding box.center)}]
        \Vertex[x = 0.2, y = 0.5, size = 0.7, style = {color=white}]{O1}
        \Vertex[label = $A^{[1]}$,size = 0.7]{A} \Vertex[x=1,label = $A^{[2]}$,size = 0.7]{C} \Vertex[y=-1,style = {color=white}]{D} \Vertex[y=-1,x = 1,style = {color=white}]{B} \Vertex[x=2,style = {color=white}]{F}
        \Edge(A)(C)
        \Edge(A)(D)
        \Edge(C)(B)
        \Edge(C)(F)
    \end{tikzpicture} \ldots \begin{tikzpicture}[baseline={([yshift=1ex]current bounding box.center)}]
        \Vertex[x = -0.2, y = 0.5, size = 0.7, style = {color=white}]{O2}
        \Vertex[label = $A^{[N]}$,size = 0.7]{A} \Vertex[x=-1,style = {color=white}]{C} \Vertex[y=-1,style = {color=white}]{D}
        \Edge(A)(C)
        \Edge(A)(D)
    \end{tikzpicture}.
\end{split}
\end{equation}

This description of an MPS is not unique as a result of gauge freedom: transformations of the type $A^{[i]}\to A^{[i]}X$, $A^{[i+1]}\to X^{-1}A^{[i+1]}$ alter the tensors used to define the MPS, but not the final state obtained on the physical legs. This problem is partially remedied by considering a restriction on the possible tensors $A^{[i]}$, which is to assume they are derived from a larger $Dd \times Dd$ unitary as
\begin{equation}
\label{eq:AfromU}
    \begin{tikzpicture}[baseline={([yshift=1ex]current bounding box.center)}]
        \Vertex[label = $A^{[i]}$,size = 0.7]{A} \Vertex[x = 1,style = {color=white}]{B} \Vertex[x=-1,style = {color=white}]{C} \Vertex[y=-1,style = {color=white}]{D}
        \Edge(A)(B)
        \Edge(A)(C)
        \Edge(A)(D)
    \end{tikzpicture} = 
    \begin{tikzpicture}[baseline={([yshift=1ex]current bounding box.center)}]
        \Vertex[label = $U^{[i]}$,shape = rectangle,size = 0.7]{A} \Vertex[x = 1,style = {color=white}]{B} \Vertex[x=-1,style = {color=white}]{C} \Vertex[y=-1,style = {color=white}]{D} \Vertex[y=0.9, label = 0, opacity =0,size=0.4]{F}
        \Edge[Direct](B)(A)
        \Edge[Direct](A)(C)
        \Edge[Direct](A)(D)
        \Edge[Direct](F)(A)
    \end{tikzpicture}.
\end{equation}

This reduces the gauge freedom to only unitary matrices $X$, and makes the MPS automatically left-canonical. This is convenient because the correlations of such tensors can be described via the right environment denoted $\Gamma_i$, which is constructed iteratively starting from the right edge of the chain according to
\begin{equation}
\label{eq:rightenv}
    \begin{tikzpicture}[baseline={([yshift=-0.6ex]current bounding box.center)}]
        \Vertex[label = $A^{[i]}$,size = 0.7]{A} \Vertex[x=-1,style = {color=white}]{C} \Vertex[y=-1, label = $\overline{A}^{[i]}$,size = 0.7]{D} \Vertex[x=-1,y=-1,style = {color=white}]{F} \Vertex[x=0.8,y=-0.5,size = 0.7,label = $\Gamma_i$,RGB,color={253,192,134}]{H}
        \Edge(A)(D)
        \Edge(A)(C)
        \Edge(D)(F)
        \Edge[bend=45](A)(H)
        \Edge[bend=45](H)(D)
    \end{tikzpicture} = \begin{tikzpicture}[baseline={([yshift=-0.6ex]current bounding box.center)}]
        \Vertex[x=-1,size = 0.7,style = {color=white}]{A} \Vertex[x=-1,y=-1, size = 0.7,style = {color=white}]{D} \Vertex[x=-0.2,y=-0.5,size = 0.7,label = $\Gamma_{i-1}$,RGB,color={253,192,134}]{H}
        \Edge[bend=45](A)(H)
        \Edge[bend=45](H)(D)
    \end{tikzpicture}.
\end{equation}
\textit{MPS Measure}---A very common measure over the space of MPS is constructed by taking the product of the invariant Haar measures of the unitaries $U^{[i]}$
\begin{equation}
    d\mu_{RMPS}(\Psi) \propto dU^{[1]}dU^{[2]}\ldots dU^{[N]},
\end{equation}
and the associated random variable was dubbed a Random Matrix Product State (RMPS).

While this is a valid measure, it is not a priori clear whether this reproduces the Fubini-Study measure, which is obtained from the determinant of the metric given by
\begin{equation}
    ds^2 = \norm{d\ket{\Psi[x]}}^2,
\end{equation}
with $d\ket{\Psi[x]}$ a vector in the local tangent space of the matrix product state manifold, parameterized by some variable $x$. A principal difficulty in checking the relation between the two measures is the need for a parameterization of the manifold that removes all gauge freedom. To achieve this, we further restrict the form of the site unitaries $U$ to those obtained via the exponential parameterization
\begin{equation}
    U(x) = U_0 \exp\left(\begin{bmatrix}
        0_D & -x^\dagger \\
        x & 0_{(d-1)D}
    \end{bmatrix}\right),
\end{equation}
in terms of the arbitrary complex $(d-1)D \times D$ matrix of coordinates $x$. $U_0$ is an arbitrary reference unitary. This completely fixes the gauge and globally recovers the entire MPS manifold \cite{Wouters2013}. Since the mapping is performed on-site and the parameterization is surjective, it must be that we can find a way to express the previously defined measure as
\begin{equation}
    d\mu_{RMPS}(x) = \prod_i J(x_i,\overline{x}_i)dx_i,
\end{equation}
with $dx$ the Lebesgue measure over the complex coordinate matrix.

Using standard results from random matrix theory \cite{Mehta2004} we prove the following expression for the Jacobian of the transformation
\begin{equation}
J(x,\overline{x}) = \abs{\frac{\sin^2\hat{x}}{\hat{x}^2}}^{D(d-2)} \abs{\frac{\sin 2\hat{x}}{2\hat{x}}}\frac{\Delta^2(\sin^2\hat{x})}{\Delta^2(\hat{x})},
\end{equation}
where $\hat{x}^2 = x^\dagger x$ is a positive hermitian matrix and $\Delta(\hat{x})$ is the Vandermonde determinant of the eigenvalues of $\hat{x}$. The details of the proof are given in the Supplementary Material.

Let us now consider the Fubini-Study measure. Since we have an explicit parameterization in the vicinity of some reference MPS, we can construct the metric at the origin by looking at small deviations
\begin{equation}
    ds^2 = dx_\nu^{(j)}d\overline{x}^{(i)}_\mu\bra{\partial_\mu^{(i)}\Psi}\ket{\partial_\nu^{(j)}\Psi} = g^{(i,j)}_{\overline{\mu}\nu} dx_\nu^{(j)}d\overline{x}^{(i)}_\mu,
\end{equation}
where $\mu$ is an index running over all entries in matrix $x$. Standard MPS contraction calculations show that the overlap is $0$ when $i \neq j$ and depends only on the right environment $\Gamma_i$ when the derivatives are placed on the same site. Carefully considering the MPS contractions that define the local distances on the manifold leads to the following volume factor associated to the determinant of the metric defined above
\begin{equation}
    \abs{g} = \prod_i \abs{\Gamma_i}^{D(d-1)},
\end{equation}
where the right environments are computed according to the reference MPS (at $x=0$). Then, we arrive at the following expression for the volume element of the Fubini-Study metric at the origin
\begin{equation}
    d\mu_{FS}(x=0) = \prod_i dx_i \abs{\Gamma_i}^{D(d-1)},
\end{equation}
which makes it a factor of $\abs{g}$ larger than the previously derived $\mu_{RMPS}$ at the origin. However, since we could have chosen any point as the reference for our parameterization, we conclude that the relation between the two measures must in fact hold everywhere on the manifold, with the condition that we compute the volume factor $\abs{g}$ at that point. This leads to the main result of this work: the typically used $\mu_{RMPS}$ measure is related to the unbiased Fubini-Study measure via the relation
\begin{equation}
    d\mu_{FS} = d\mu_{RMPS} \prod_i \abs{\Gamma_i}^{D(d-1)}.
\end{equation}
It is worth noting that the additional factor only depends on the eigenvalues of the right environment matrices $\Gamma_i$, which are directly related to the Schmidt coefficients of a bipartition of the chain between sites $i$ and $i+1$. Being purely a function of the entanglement structure of the MPS, it is invariant both under gauge transformations and single-site unitaries applied to the physical legs.

\textit{Properties}---Here, we discuss some properties relating to the newly introduced distribution. An important construction for many use cases is having a resolution of identity over the semi-classical phase space of matrix product states. We would then like to consider the result of the integral
\begin{equation}
    \int d\mu_{FS}(\Psi) \ket{\Psi}\bra{\Psi}.
\end{equation}

The key is to realize that the measure is invariant under local unitary rotations on each physical site. If we consider a Haar random local unitary $V_i$ applied to each site the integral can be equivalently states as
\begin{equation}
    \int d\mu_{FS}(\Psi) \prod_i dV_i \bigotimes_i V_i\ket{\Psi}\bra{\Psi} \bigotimes_i V_i^\dagger.
\end{equation}

Using standard results in Haar integration to perform the integrals over $V_i$ we find that
\begin{equation}
    \frac{1}{d^N} \int d\mu_{FS} (\Psi) \Tr(\ket{\Psi}\bra{\Psi}) I_{d^N} = I_{d^N}\int d\mu_{FS}(\Psi),
\end{equation}
which is proportional to the many-body identity $I_{d^N}$ over the chain, with a coefficient given by the following partition function over random unitaries measuring the size of the manifold under the new measure
\begin{equation}
    \mathcal{Z} = \int \prod_i dU_i e^{-D(d-1)\sum_i \log \abs{\Gamma_i}^{-1}}.
\end{equation}

We also consider the entanglement properties of samples drawn from $\mu_{FS}$. At first glance, the effect of the determinant factor in the measure is to repel the eigenvalues of the right-environment away from the origin, which can reasonably be expected to give rise to an increase in entanglement entropy across the chain. In the Supplementary Material, we show that the partition function above can be obtained from a transfer matrix acting in the space of probability distributions over the right environments $\Gamma_i$. It is dominated by the distribution corresponding to its largest eigenfunction, which sets the typical entanglement spectrum across the chain.
If one ignores the determinant factor and works with the partition function of the RMPS ensemble instead, the typical spectrum (for sufficiently large D) is concentrated around the famous Marchenko-Pastur (MP) distribution with an aspect ratio of $c = 1/d$ \cite{Marchenko1967}. This is the eigenvalue distribution of a complex Wishart matrix with $p = D$ features and $n = dD$ samples. In a first-order approximation, we show that moving to the FS ensemble corresponds to changing the number of samples of the Wishart matrix to $n' = D(2d-1)$, or equivalently modifying the aspect ratio to the new value $c' = 1/(2d-1)$. This is not a self-consistent solution and the actual distribution obtained from sampling seems to deviate from this, although the agreement is surprisingly good. More details about this can be found in the Supplementary Material.

\begin{figure}[t!]
    \centering
    \includegraphics[width=0.47\textwidth]{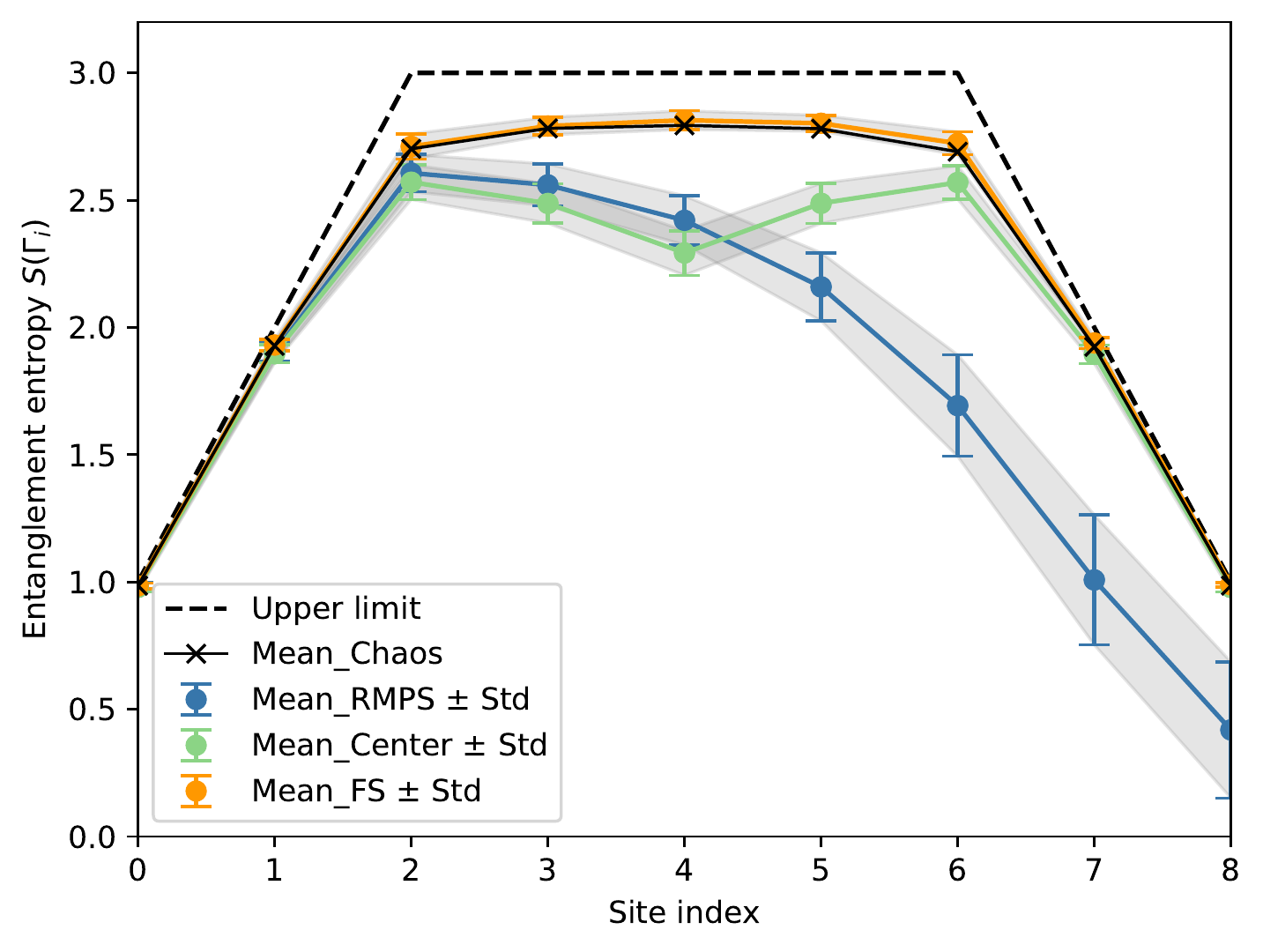}
    \caption{Plot of the entanglement entropy across a length $N = 10$ qubit chain in a nearest-neighbor chaotic quantum system following TDVP (black crosses), the left canonical RMPS, central gauge RMPS and FS ensembles, at maximum bond dimension $D=8$. Means and standard deviations are based on batches of 200 independent samples. The dashed black line marks the maximal entanglement allowed by the manifold.}
    \label{fig:ent_chain}
\end{figure}

\begin{figure}[t!]
    \centering
    \includegraphics[width=0.47\textwidth]{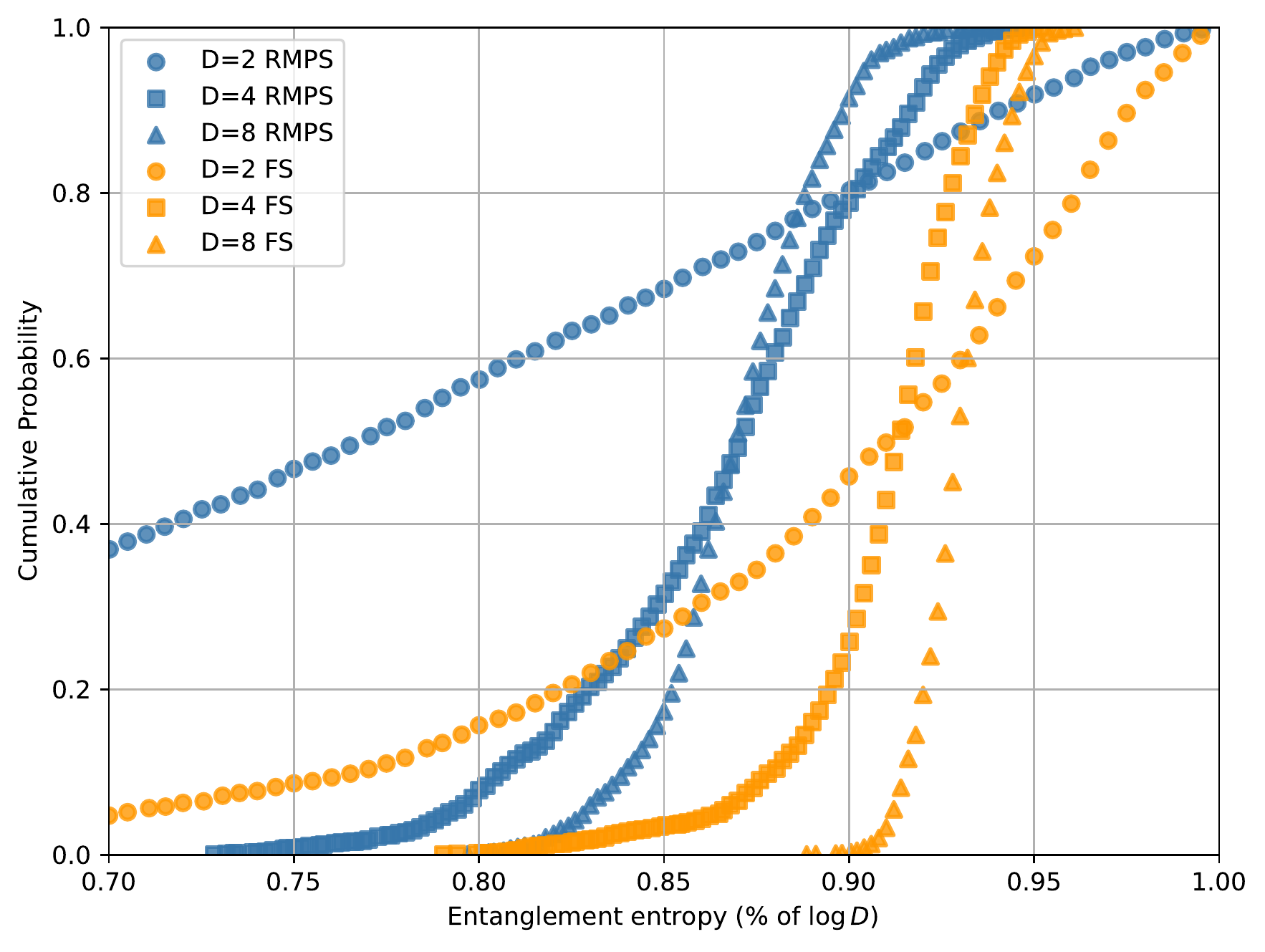}
    \caption{Cumulative distribution function of the entanglement entropy sampled from the equilibrium right-environment distribution, plotted as a function of percentage of the maximum entanglement allowed by the manifold $\log_2 D$. All plots are taken using a chain of $N=10$ qubits $d=2$.}
    \label{fig:ent_hist}
\end{figure}

\begin{figure}[t!]
    \centering
    \includegraphics[width=0.47\textwidth]{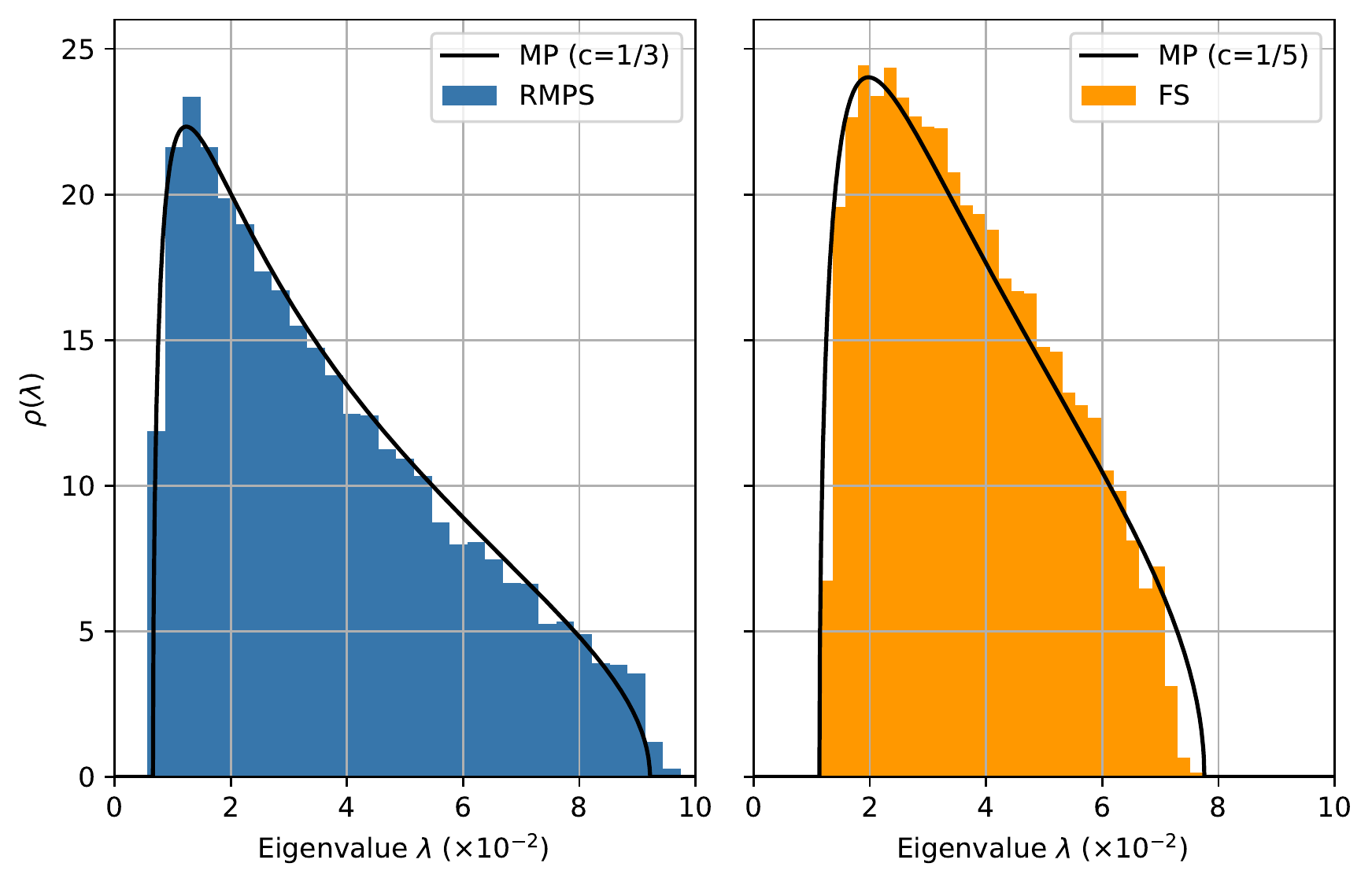}
    \caption{Eigenvalue histogram of 1000 random right-environment matrices sampled from the equilibrium distribution, for a system with $N=10$ sites, local dimension $d=3$ and bond dimension $D=27$. Black lines show the theoretically predicted Marchenko-Pastur distributions at aspect ratio $c = 1/d$ and $c=1/(2d-1)$ for the RMPS and FS distributions respectively.}
    \label{fig:eigv_hist}
\end{figure}

\textit{Sampling}---In this section we discuss a straight-forward method for sampling from the FS distribution, based on an adaptation of the Metropolis-Hastings algorithm.

We start with an initial matrix product state from the standard RMPS ensemble, produced by sampling independent Haar random unitaries at each site. We then sequentially pass through the chain and propose small fluctuations to the on-site unitaries. If a modification is proposed at site $j$, the acceptance ratio is given by
\begin{equation}
    \alpha = \frac{\prod_{i<j}\abs{\Gamma'_i}^{D(d-1)}}{\prod_{i<j} \abs{\Gamma_i}^{D(d-1)}},
\end{equation}
noting that a change in the unitary at site $j$ will only affect the right environments of the sites to its left. We then generate a uniform random number $u$ in the interval $[0,1]$ and accept the modification if $u < \alpha$. The sweeping through the chain is performed until an equilibrium distribution is reached. We see rapid mixing in the numerical simulations we performed, though the scaling of the burn-in time with bond dimension and the size of the chain is currently unknown.

In Fig.~\ref{fig:ent_chain} we show the entanglement profile of a chain of $N = 10$ qubits in both the $RMPS$ and $FS$ ensembles of a $D=8$ MPS manifold. It is immediately obvious that the statistical properties of entanglement in the $RMPS$ ensemble are not invariant under a flip, due to the choice of constructing the measure based on the left-canonical MPS parameterization. One could propose mitigating this by moving the center of orthogonality to the middle of the chain before sampling the unitaries, but we find that this produces a discontinuity in the entanglement entropy. The $FS$ measure does not suffer from either of these issues. We also notice that the edge effects are more pronounced in the $RMPS$ ensemble, such that it takes longer for the right environment to converge to its equilibrium distribution. For comparison, we also performed a time-dependent variational principle (TDVP) simulation of the state of a chaotic nearest-neighbor quantum Hamiltonian starting from a product state. The Hamiltonian includes ferromagnetic couplings $J_x =1$, $J_y = 0.9$ and local fields $h_z = 0.1$, $h_x = 0.8$, $h_y = 0.3$, chosen arbitrarily. TDVP is known to conserve energy, so we do not expect an ergodic exploration of the MPS manifold in this case. Despite this, we see that the entanglement statistics very closely resemble those of the FS ensemble, providing additional evidence for its generality.

The entanglement histogram of the converged distribution of the right-environment is shown in Fig.~\ref{fig:ent_hist}. The repulsion of eigenvalues away from the origin induced by the determinant factors indeed results in a larger average entanglement.

Finally in Fig.~\ref{fig:eigv_hist} we show a histogram of the eigenvalues of matrices sampled from the equilibrium right environments in the $RMPS$ and $FS$ distributions. The $RMPS$ case is well approximated by an MP distribution with $c = 1/d$, corroborating the theoretical prediction. For the $FS$ eigenvalues, the approximation by an MP distribution is less accurate, with some deviation visible at the higher end of the spectrum.

\textit{Discussion}---Our work identifies a canonical ensemble of matrix product states, aligned with the typical Fubini-Study measure on the quantum Hilbert space. Previous heuristic constructions, such as placing Haar-random unitary matrices at each site, do not produce an unbiased sampling over the MPS manifold, leading to spurious left-right asymmetry and modified entanglement statistics. By deriving the proper volume element, this work establishes a canonical definition for a truly random MPS, ensuring uniform coverage throughout the Hilbert space and eliminating arbitrary gauge dependence. This provides a robust foundation for statistical analysis of tensor networks, in line with well-established notions of randomness in full Hilbert space.

We expect the uniform measure will have broad implications for quantum many-body physics, particularly in studies of entanglement structure and thermalization, ensuring the results reflect genuine typicality rather than artifacts of the sampling procedure. We anticipate the large bond dimension limit of this ensemble may provide a way forward in both analytical and numerical studies of entanglement phase transitions \cite{Cecile2024}, and highly correlated quantum magnetism \cite{Green2016, Leontica2024}. Equipped with a homogeneous distribution over a semi-classical phase space and the time-dependent variational principle as a method of generating classical mechanics for arbitrary many-body systems, we can employ the machinery of statistical mechanics to study entanglement dynamics in a fully classical setting. This will be instrumental in establishing the projected Lyapunov spectrum as a characterization tool for many-body quantum chaos \cite{Leontica2025, Hallam2019}.

Similar concerns over the use of statistically independent tensors apply for more general networks, such as PEPS, tree tensor networks or MERA. For MPS, the connection between the product of local measures and the globally unbiased measure is directly related to the entanglement spectrum of bipartitions, so it would be interesting to check how this result generalizes to more complex networks. Tree tensor networks would be a feasible candidate, due to the absence of loops, but this is left for future work.

The sampling technique, based on the Metropolis-Hastings algorithm, ensures that these subroutines can be implemented in practice, although sequential sampling alternatives may be required to make this approach more efficient. We believe this should be possible and will be the subject of future work.

Overall, the work establishes a canonical definition for random MPS that will likely play a foundational role for future research and applications.

\vspace{0.5cm}
\textit{Acknowledgments}---S.L. acknowledges support from UCL’s Graduate Research Scholarship and Overseas Research Scholarship.

\bibliography{bibliography}

\onecolumngrid


\section{Jacobian of exponential parameterization}

In this section we consider the Jacobian of the mapping from the $(d-1)D \times D$ complex matrix $x$ to the on-site unitaries $U(x)$ under the exponential mapping given in the main text
\begin{equation}
    U(x) = \exp\left(\begin{bmatrix}
        0_D & -x^\dagger \\
        x & 0_{(d-1)D}
    \end{bmatrix}\right),
\end{equation}
where we chose the reference $U_0 = I$ for convenience, as this does not affect the measure.

Since the parameterization does not cover all unitaries, but only a gauge-fixed subset, our task is to find a function $J(x,\overline{x})$ such that integrals of gauge-invariant functions can be rewritten as
\begin{equation}
    \int \prod_i dU_i f(\Psi) = \frac{1}{Z}\int_D \prod_i dx_i J(x_i,\overline{x}_i) f(\Psi),
\end{equation}
where $\Psi$ is the MPS constructed from the unitaries, $dx$ is the Lebesgue measure over the complex matrix $x$ and $Z$ is a normalization prefactor independent of the integrand $f$. 

The first step is to note that we can decompose
\begin{equation}
    U = \begin{bmatrix}
        A & B \\
        C & D
    \end{bmatrix},
\end{equation}
and only $A$ and $C$ play a role in the construction of $\Psi$. Integrating out the unitary completion, we arrive at the conclusion
\begin{equation}
    dU \propto dA dC \delta (A^\dagger A + C^\dagger C - I).
\end{equation}

Using the polar decompositions $A = U_A \sqrt{W_A}$ and $C = U_C \sqrt{W_C}$, with the associated measure transformations derived from the theory of complex Wishart matrices, we can further rewrite this as
\begin{equation}
\begin{split}
    dU &\propto dU_A dU_C dW_A dW_C \abs{W_C}^{(d-2)D}   \\
    &\times\chi(W_A)\chi(W_C) \delta (W_A + W_C - I) \\
    &\propto dU_A dU_C dW \abs{W}^{(d-2)D} \chi(W)\chi(I-W),
\end{split}
\end{equation}
where $\chi(\cdot)$ imposes the positivity constraint. In terms of these parameters we have $A = U_A \sqrt{I-W}$ and $C = U_C \sqrt{W}$. With this formulation we see that the correct value of $x$ that recovers $A$ and $C$ under exponentiation is
\begin{equation}
    x = U_C \arcsin \sqrt{W} U_A^\dagger.
\end{equation}

The Lebesgue measure over $x$ can be obtained from the following measure over $U_C$, $U_A$ and $W$
\begin{equation}
    \begin{split}
        dx &\propto dU_C dU_A \abs{f(W)}^{(d-2)D}\abs{\sqrt{\frac{f(W)}{W(I-W)}}} \\
        &\times \frac{\Delta^2(f(W))}{\Delta^2(W)} dW,
    \end{split}
\end{equation}
where $f(W) = (\arcsin(\sqrt{W}))^2$.

Comparing this expression to the one of the local unitary matrices above we get the following relation

\begin{equation}
\begin{split}
    dU \propto dx \abs{\frac{W}{f(W)}}^{(d-2)D}\abs{\sqrt{\frac{W(I-W)}{f(W)}}}\frac{\Delta^2(W)}{\Delta^2(f(W))}.
\end{split}
\end{equation}
We now wish to express the Jacobian in terms of the singular values of $x$, or equivalently in terms of the eigenvalues of $\hat{x} =\sqrt{x^\dagger x}$. Using this we have

\begin{equation}
    dU \propto dx \abs{\frac{\sin^2 \hat{x}}{\hat{x}^2}}^{D(d-2)} \abs{\frac{\sin(2\hat{x})}{2\hat{x}}} \frac{\Delta^2(\sin^2 \hat{x})}{\Delta^2(\hat{x}^2)}
\end{equation}

To first order, the Jacobian $J(\hat{x})$ is approximately given by
\begin{equation}
    J(\hat{x}) \approx 1- \frac{Dd}{3}\Tr \hat{x}^2 \approx e^{-\frac{Dd}{3}\Tr\hat{x}^2}.
\end{equation}

\section{Right environment dynamics}

The measure we propose comes with its own statistics that differ significantly from the RMPS ensemble. In this section, we focus on the partition function 
\begin{equation}
    \mathcal{Z} = \int \prod_i dU_i \abs{\Gamma_i}^{D(d-1)},
\end{equation}
introduced in the main text and use it to derive an approximate form for the spectrum of the right environment, which governs entanglement across a cut. Since the right environments obey the recurrence relation
\begin{equation}
    \Gamma_{i-1} = \Tr_p{U_i \Gamma_{i} \otimes P_0 U_i^\dagger},
\end{equation}
we can equivalently express the partition function in a more local form as
\begin{equation}
    \mathcal{Z} = \int \prod_i dU_i d\Gamma_i \abs{\Gamma_i}^{D(d-1)} \delta(\Gamma_{i-1}-\Tr_p{U_i \Gamma_{i} \otimes P_0 U_i^\dagger}),
\end{equation}
or more compactly as
\begin{equation}
    \mathcal{Z} = \int \prod_i d\Gamma_i \mathcal{K}(\Gamma_{i-1}, \Gamma_i),
\end{equation}
where $\mathcal{K}$ is given by
\begin{equation}
    \mathcal{K}(\Gamma_{i-1}, \Gamma_i) = \int dU \abs{\Gamma_{i-1}}^{D(d-1)} \delta(\Gamma_{i-1}-\Tr_p{U \Gamma_{i} \otimes P_0 U^\dagger}).
\end{equation}

This form makes it clear that the many-body partition function can be computed via a transfer matrix approach, albeit in the complex space of probability distributions over the right environments. The linear transfer functional is given via the integral form
\begin{equation}
    T[P](K) = \int d\Gamma \mathcal{K}(K,\Gamma) P(\Gamma),
\end{equation}
which allows us to write the partition function symbolically as
\begin{equation}
    \mathcal{Z} = \bra{l} T^N \ket{r},
\end{equation}
with $l$ and $r$ some boundary conditions that do not affect the typicality properties of the entanglement spectrum for sufficiently large N. Assuming $T$ has a finite correlation length $\xi$ and a non-degenerate leading eigenvalue $\lambda$, we can write an approximate expression for $T^N$ at large N as
\begin{equation}
    T^N  = \lambda^N\left( \ket{P_\lambda}\bra{\phi_{\lambda}} + \mathcal{O}(e^{-N/\xi})\right),
\end{equation}
where the leading left and right eigenfunctions are the solutions of the integral equations
\begin{align}
    \int dK \phi_\lambda(K) \mathcal{K}(K,\Gamma) = \lambda \phi_\lambda(\Gamma),\\
    \int d\Gamma P_\lambda(\Gamma) \mathcal{K}(K,\Gamma) = \lambda P_\lambda(K).
\end{align}

To gain some insight into the meaning of these equations, let us first consider the simplified case of the $RMPS$ ensemble, for which the kernel $\mathcal{K}^0$ has the simpler form
\begin{equation}
    \mathcal{K}^0(K,\Gamma) = \int dU \delta(K-\Tr_p{U \Gamma \otimes P_0 U^\dagger}),
\end{equation}
representing a Markov chain over the space of positive matrices. Since this is known to preserve the normalization of probability distributions we immediately have that $\lambda = 1$ and the left eigenfunction is $\phi_\lambda (K) = 1$, the uniform distribution. Using free probability theory, one can also check that the fixed point of the transformation $\Gamma \to \Tr_p{U \Gamma \otimes P_0 U^\dagger}$ in the large $D$ limit will have the eigenvalue distribution of the complex Wishart ensemble with aspect ratio $1/d$
\begin{equation}
    p_W(\Gamma) \propto \abs{\Gamma}^{n-p} \delta(\Tr\Gamma - 1),
\end{equation}
with $n = dD$ and $p = D$. In the following subsection we use techniques from free probability to prove this.

\subsection*{Limiting spectrum of right environment in RMPS}

For simplicity we will assume here that the right environments are normalized such that $\Tr \Gamma = D$. Since all transformations are linear, the case of interest can be recovered from the final result by a rescaling of the entire spectrum. We use this convention so the moments
\begin{equation}
    M_n = \frac{1}{D} \int d\Gamma P(\Gamma)\Tr \Gamma^n,
\end{equation}
remain finite in the thermodynamic limit $D \to \infty$. Under the effect of the transformation introduced above, the moments evolve to
\begin{equation}
    M_n' = \frac{1}{D}\int dU d\Gamma P(\Gamma) \Tr\left[T(\Gamma)^n\right],
\end{equation}
where $T(\Gamma) = \Tr_p\left(U \Gamma\otimes P_0 U^\dagger\right)$. If we introduce an additional system of dimension $d$, we can remove the trace over the physical leg in the expression above to leave only matrix multiplications
\begin{equation}
    T(\Gamma) = I\otimes \bra{\Psi^+} \left(U\Gamma\otimes P_0 U^\dagger\right) I\otimes \ket{\Psi^+},
\end{equation}
where we use the notation
\begin{equation}
    \ket{\Psi^+} = \sum_{i=0}^{d-1} \ket{i}\otimes\ket{i},
\end{equation}
for the vectorization of the identity into the doubled physical space.

Its moments can be equivalently written in terms of the 2 hermitian matrices $A = I\otimes \ket{\Psi^+}\bra{\Psi^+}$ and $B = (U\Gamma\otimes P_0 U^\dagger) \otimes I$. We can show that their $D \to \infty$ limits form freely independent random variables. We can prove this directly using the definition of free independence, which is that traces of mixed products of centered random matrices vanish in the large $D$ limit, or

\begin{equation}
    \frac{1}{d^2 D}\Tr\left((B^{n_1}-\frac{1}{d^2D}\Tr B^{n_1})(A-\frac{1}{d})(B^{n_2}-\frac{1}{d^2D}\Tr B^{n_2})\ldots\right) \to 0,
\end{equation}
where only the first power of $A$ needs to be considered as it is proportional to a projector with a prefactor independent of $D$. It can be easily checked that the centered matrix $A - 1/d$ for $d = 2$ can be rewritten as
\begin{equation}
    A-\frac{1}{2} =\frac{1}{2} I\otimes(X\otimes X - Y\otimes Y+Z\otimes Z),
\end{equation}
in terms of the Pauli matrices $X, Y, Z$. Then one can see that to prove the above it is sufficient to prove
\begin{equation}
    \frac{1}{2 D}\Tr\left((\tilde{\Gamma}^{n_1}-\frac{1}{2 D}\Tr \tilde{\Gamma}^{n_1})I\otimes P_1(\tilde{\Gamma}^{n_2}-\frac{1}{2 D}\Tr \tilde{\Gamma}^{n_2})I\otimes P_2\ldots\right) \to 0,
\end{equation}
for all choices of the Pauli operators $P_1,P_2\ldots$, where $\tilde{\Gamma} = U\Gamma \otimes P_0 U^\dagger$. This is because, upon expanding the factors of $A-1/2$, each resulting term in the sum will contain such a factor multiplied by an $O(1)$ term coming from the contraction on the auxiliary physical space. We know the above holds, because $\tilde{\Gamma}$ acts in a Haar random basis with respect to the deterministic set of matrices $I\otimes P$. This guarantees that the subalgebras generated by $\tilde{\Gamma}$ and the Paulis $I\otimes P$ are freely independent at large $D$. The proof can be easily extended to arbitrary physical dimension $d$ by noting that one can always perform an SVD of $\ket{\Psi^+}\bra{\Psi^+}$ that decouples the 2 physical spaces, then proceed similarly.

Then the problem of finding $M_n'$ given $M_n$s is equivalent to finding the moments of the product $AB$ of 2 freely independent variables. This is a well-studied problem in free probability theory and the solution is given as a relation between $S$-transforms
\begin{equation}
    S_{AB}(z) = S_A(z)S_B(z),
\end{equation}
where $A,B$ are free random variables. Considering the series of moments
\begin{equation}
    M_A(z) = \sum_{n=1}^\infty M_n^{(A)} z^n,
\end{equation}
the S-transform is defined by
\begin{equation}
    S_A(z) = \frac{1+z}{z}M_A^{\langle-1\rangle}(z),
\end{equation}
where $\langle-1\rangle$ denotes the inverse under composition. First, we want to consider the S-transform of the known matrix A. Given that $M_n^{A} = \Tr A^n /(Dd^2) = d^{n-2}$ we have
\begin{align}
    M_A(z) &= \frac{1}{d^2} \frac{zd}{1-zd}, \\
    M_A^{\langle-1\rangle}(z) &= \frac{1}{d}\frac{zd^2}{1+zd^2}, \\
    S_A(z) &= d \frac{1+z}{1+d^2z}.
\end{align}

We also need to relate the S-transform of $B$ to the S-transform of $\Gamma$. Following their relation we find that
\begin{equation}
    S_{B}(z) = \frac{d(1+z)}{1+zd} S_\Gamma (dz).
\end{equation}

Putting the equations together we find that the S-transform evolves according to
\begin{equation}
    S_{AB}(z) = d\frac{1+z}{1+d^2 z} \frac{d(1+z)}{1+zd} S(dz).
\end{equation}

Following a similar calculation as done previously we find that the S-transforms of $AB$ and $T(\Gamma)$ are related via
\begin{equation}
    S_{T(\Gamma)}(zd^2) = \frac{1+zd^2}{d^2(1+z)} S_{AB}(z),
\end{equation}
giving us the final equation for propagating the S-transforms of the right environments
\begin{equation}
    S'(z) = \frac{1+\frac{z}{d^2}}{1+\frac{z}{d}}S(\frac{z}{d}).
\end{equation}

A fixed point of the above is given by
\begin{equation}
    S(z) = \frac{1}{1+\frac{z}{d}},
\end{equation}
which is the known S-transform of the Marchenko-Pastur distribution with aspect ratio $c = 1/d$.

This agrees very well with numerical simulations, as seen in the main text. 

\subsection*{Approximate spectrum of FS-MPS}

We can use the known distribution of the RMPS right environment to obtain a first order approximation of the stationary distribution for the full transfer matrix. We see that passing this distribution through the transfer matrix of the FS partition function we get
\begin{equation}
    p_W'(K) \propto \abs{K}^{D(d-1)} p_W(K) \propto \abs{K}^{n'-p} \delta(\Tr K - 1),
\end{equation}
where $p=D$ is again the dimension of the matrix but $n' = D(2d-1)$, suggesting a Wishart distribution with the aspect ratio changed to $1/(2d-1)$. This is a very crude approximation of the typical structure of the right environment, as it is not exactly stationary and it ignores the contribution to the probability distribution due to the transformed left eigenfunction. Nevertheless, the agreement with numerical results is good, though some distortions are visible, particularly near the top of the spectrum. This shows that more work is necessary to determine the exact shape of the spectrum.

\end{document}